\documentclass[paper,nofootinbib] {revtex4}
\pdfoutput=1
\usepackage{graphicx}
\usepackage{epsfig}
\usepackage{fancyhdr}
\usepackage{empheq}
\usepackage{mathbbol}
\usepackage{wasysym}
\usepackage{pstricks}
\usepackage{color}
\usepackage{bbm}
\newcommand{\be}{\begin{equation}}
\newcommand{\ee}{\end{equation}}
\newcommand{\bea}{\begin{eqnarray}}
\newcommand{\eea}{\end{eqnarray}}

\begin{document}

\title{Probing Minimal Supersymmetry at the LHC with the Higgs Boson Masses}
%
\author{L. Maiani}
\affiliation{Department of Physics and INFN, `Sapienza' Universit\`a di Roma,
Piazzale A. Moro 5, Roma, I-00185, Italy
}
\author{A. D. Polosa}
\affiliation{Department of Physics and INFN, `Sapienza' Universit\`a di Roma,
Piazzale A. Moro 5, Roma, I-00185, Italy
}
\author{V. Riquer}
\affiliation{Fondazione TERA, Via G. Puccini 11, I-28100, Novara, Italy
}

\begin{abstract}
ATLAS and CMS report indications of a Higgs boson at $M_h\sim125$ GeV. In addition, CMS data show a tenuous bump 
 in  the $ZZ$ channel, at about $320$~GeV. We make the bold assumption that it might be the indication of a secondary line corresponding to the heaviest scalar Higgs boson of Minimal Supersymmetry, H, and discuss the viability of this hypothesis. We discuss also the case of a heavier $H$. The relevance of the $b \bar b$ decay channel is underlined.
\end{abstract}

\maketitle

\thispagestyle{fancy}

{\bf \emph{Introduction}}.
Indications of a Higgs boson around a mass:
\be
M_h\simeq 125\;{\rm GeV } 
\label{mass}
\ee
have been presented by the ATLAS~\cite{atlas1} and CMS~\cite{cms1} collaborations in the reactions:
\bea
&&p+p\to h +{\rm All} \to \gamma \gamma +{\rm All} \label{gamgam} \\
&&p+p\to h + {\rm All} \to ZZ + {\rm All}  \;\; (ZZ \to 4 \;{\rm charged\;leptons})
\label{ZZ}
\eea
observed at the LHC at c.o.m. energy: $\sqrt{s}=7$ TeV.
 
The value of the mass, should it be confirmed, is interesting by itself, in that it is compatible with the restrictions posed by the Minimal Supersymmetric Standard Model \cite{MSSM,barger-altri}, MSSM in short. It is well known that the tree level inequality for the mass of the lightest Higgs boson:
\be
M_h^2  \leq \cos^2(2\beta) M^2_Z \;{\rm (tree\;level)}
\label{treelvl}
\ee
receives radiative corrections, due to the exchange of the top quark and its two scalar partners, $\tilde t_{L,R}$, which may bring the mass up to $M_h \sim 135$ GeV, for ``reasonable" values of the scalar partner masses.  
Encouraged by this fact, we have investigated the consequences of the next simple prediction of MSSM, namely that there must be  two Higgs doublets, one coupled to $up$ and the other to $down$ fermions. 

	The introduction of a singlet superfield~\cite{NMSSM,Ellwanger:2009dp} leads to the Next-to-Minimal Superymmetric Standard Model (NMSSM), which modifies the tree level inequality (\ref{treelvl}) thereby reducing the role of the radiative corrections and allowing for lighter scalar top quark partners and better naturalness~\cite{Hall:2011aa}. However, predictivity in NMSSM is greatly reduced for what concerns the properties of the scalar Higgs particles and we think it wise to stick to the minimal option unless some real contradiction with data is found.
	 
The mass-matrix of the two, $CP$-even, Higgs particles of the MSSM, $h$ and $H$ (with $h$ the lightest), depends upon four parameters:  the VEV ratio, parametrized in terms of an angle $\beta$, the masses of the $Z$ boson and of the $CP$-odd Higgs boson, $A$, and the average mass of the top scalar partners, which appears in the radiative correction mentioned above. Here, we take the usual notation:
\bea
&&\langle0|H^0_u|0\rangle=v \sin\beta;\;\;  \langle0|H^0_d|0\rangle=v \cos\beta; \; \; 0< \tan\beta< +\infty     \label{vevs} \\
&& v^2=(2 \sqrt{2} G_F)^{-1}=(174\;{\rm GeV})^2
\label{standardvev}
\eea 

Diagonalizing the mass matrix, we obtain two eigenvectors which express  the physical Higgs fields in terms of $H^0_{d, u}$ and determine the coupling of $h$ and $H$ to quarks, leptons and gauge bosons~\cite{susybasic,fermiofob}.

Assume now the mass in (\ref{mass}) and assume that we know the other mass, $M_H$, as well. We can express everything in terms of $M_{Z, h, H}$ and remain with one unknown parameter only, namely $\tan\beta$. We can derive all the observable quantities, such as the ratios of $\sigma \times BR$  in the MSSM to the one in the SM for the channels observed in (\ref{gamgam}) and (\ref{ZZ}), and see (i) how the level of observation of the $125$ GeV signal compares to the SM; (ii) what is the visibility level of H in channels ~(\ref{gamgam}) and (\ref{ZZ}), and (iii) which are the best suited channels for the observation of $H$. In addition, we may determine, always in terms of $\tan\beta$, the mass value of the $CP$-odd boson, $A$, and the mass scale of the top scalar partners.

{\bf \emph{Details of the calculation}}.
In the basis ($H_d, H_u$), the mass matrix of the $CP$-even Higgs fields is given by:
\bea
&&{\cal M}_S^2=M_Z^2\left(\begin{array}{cc}  \cos^2\beta & -\cos\beta\sin\beta \\ -\cos\beta\sin\beta & \sin^2\beta \end{array}\right) 
+ M_A^2\left(\begin{array}{cc}  \sin^2\beta & -\cos\beta\sin\beta \\ -\cos\beta\sin\beta & \cos^2\beta \end{array}\right) +
\left(\begin{array}{cc} 0 & 0 \\0 &\delta \end{array}\right) 
\label{massmatrix}
\eea
with $\delta$ the radiative correction:
\be
\delta = \frac{3 \sqrt{2}}{\pi^2\sin^2\beta } G_F (M_t)^4t;\nonumber \;\;\; t= \log\left(\frac{ \sqrt{M_{\tilde t_R}M_{\tilde t_L}}}{M_t}\right)
\ee
The first term in ${\cal M}_S^2$ arises from the so-called Fayet-Iliopoulos term~\cite{Fayet:1974jb} determined by the gauge interaction. In the radiative corrections we have kept only the top-stop contribution which is by far the dominating one.

We also note that in MSSM:
\be
M_{H^\pm}^2= M_A^2 + M_W^2
\label{mcharged}
\ee
 
We eliminate $M_A^2$ from (\ref{massmatrix}) equating the trace of ${\cal M}_S^2$ to the sum of the eigenvalues, $M_h^2+M_H^2$, and obtain $t$ as a function of $\tan\beta$ from the difference. One obtains two real solutions:
\be
t=F^{(\pm)}(\tan\beta)
\label{logmass}
\ee
only for 
\be
\tan\beta \geq 0.89
\ee
which therefore provides an absolute lower bound, roughly compatible with the border of the shaded exclusion region in $\tan\beta$ (for the definition of the exclusion regions, see below under {\it{\bf Results}}). 

\begin{figure}[htb]
\begin{minipage}[t]{80mm}
\includegraphics[scale=.6]{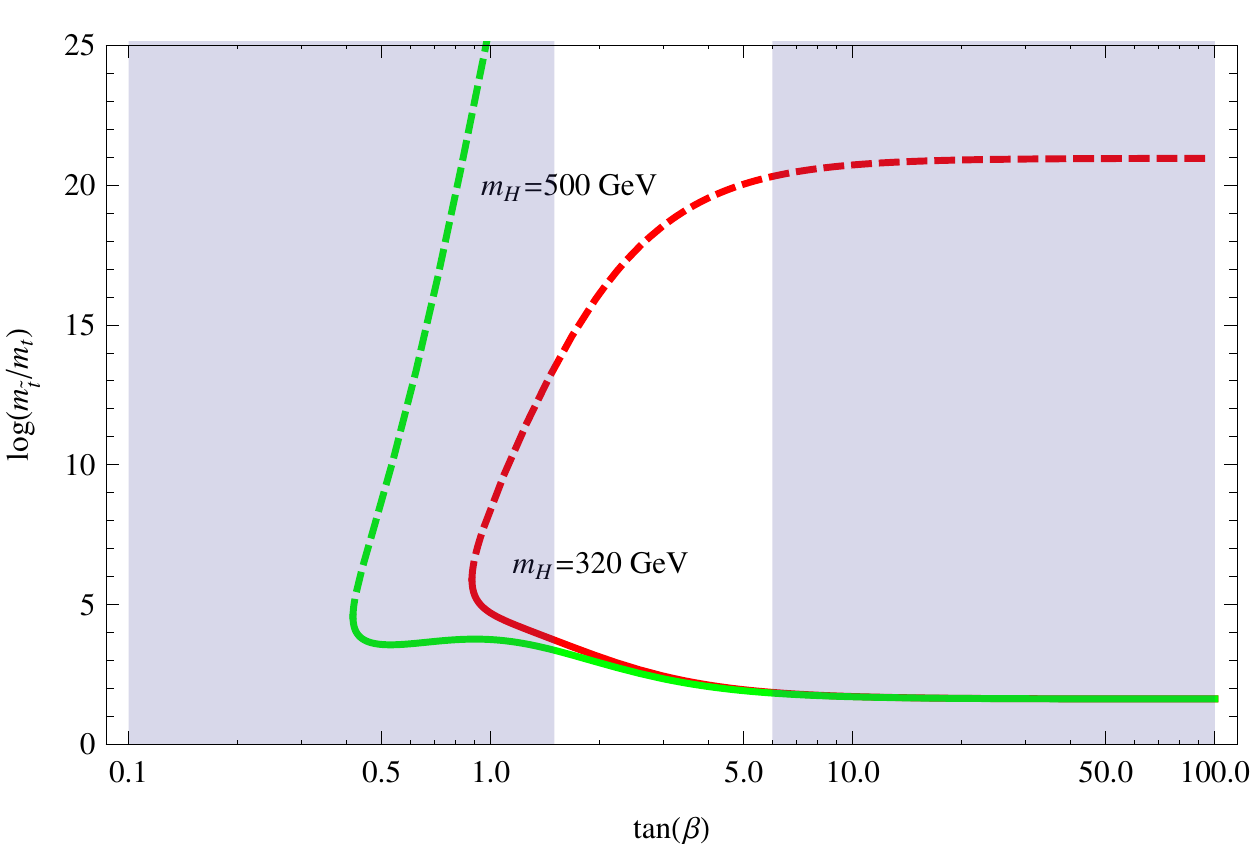}       
\caption{\label{nasoverde} \footnotesize Solutions of eq. (\ref{logmass}) vs. $\tan\beta$ for the values of $M_H$ considered in the text. }
\end{minipage}
\hspace{\fill}
\begin{minipage}[t]{75mm}
\includegraphics[scale=.6]{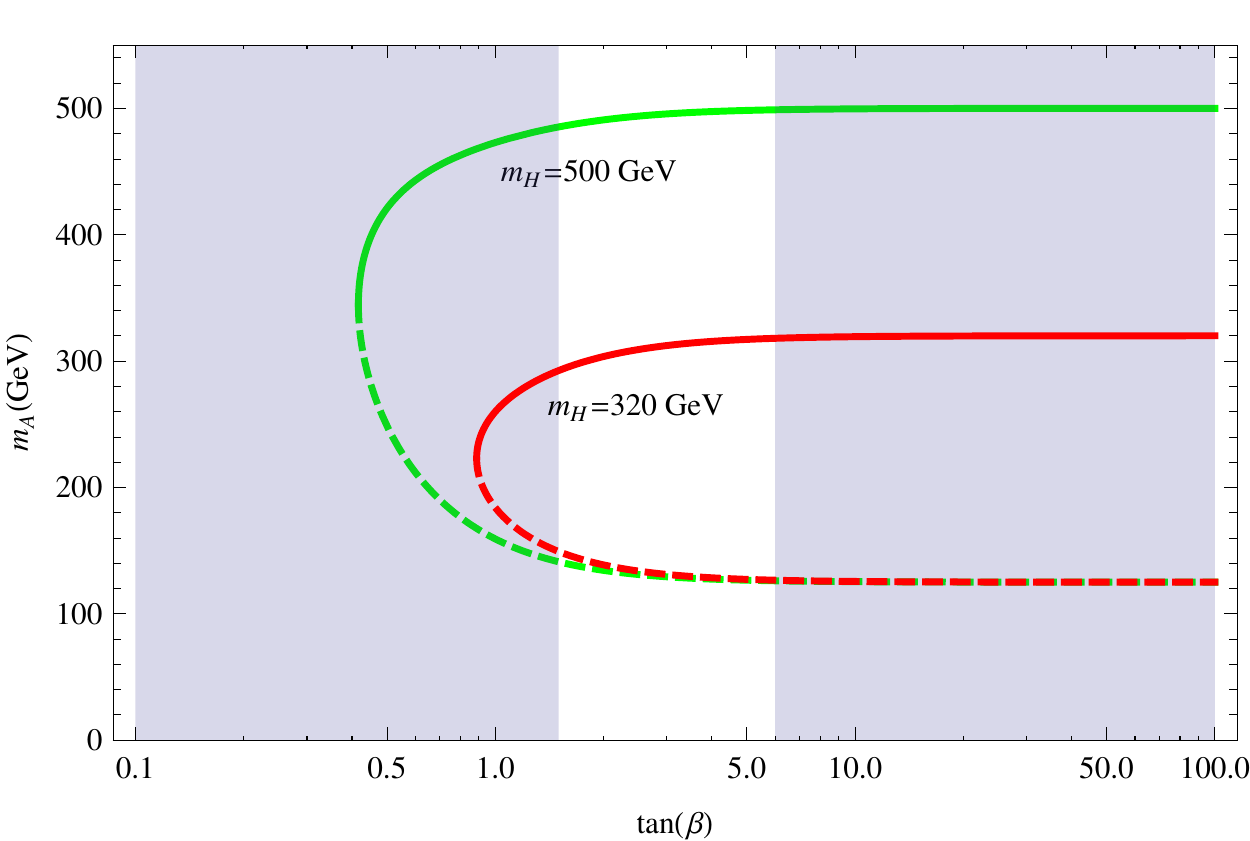}   
\caption{\label{pseudo} \footnotesize Values of the axial meson mass vs. $\tan\beta$ for the values of $M_H$ considered in the text.}
\end{minipage}
\end{figure}

The function in (\ref{logmass}) is plotted versus $\tan\beta$ in  Fig. \ref{nasoverde} for the two guiding values of $M_H$. The solid (dotted) lines correspond to $F^{(-)}$ ($F^{(+)}$). We shall choose $F^{(-)}$, which minimizes the size of the radiative correction.  In correspondence, we plot  in Fig. \ref{pseudo} the values of $M_A$, solid (dotted) lines referring to the solid (dotted) line solution in Fig.  \ref{nasoverde}.

Substituting this function back into (\ref{massmatrix}), we compute the two eigenvectors:
\bea
&&{\bf S}_h(\tan\beta)=\left(S_{hd}, S_{hu}\right)\nonumber \\
&&{\bf S}_H(\tan\beta)=\left(S_{Hd}, S_{Hu}\right)
\label{eigenvectors}
\eea
with the physical fields given by:
\be
h= S_{hi} H_i \;\;\; H=S_{Hi}H_i \;\;\; ( i=d,u)
\ee

Couplings of $h$ and $H$ to exclusive channels, e.g. $WW$, $t\bar t$, etc., are given by the SM coupling multiplied by factors which depend upon the components of the eigenvectors in (\ref{eigenvectors}) and $\beta$, see Tab. \ref{couplings}. The two components of ${\bf S}_h$ are positive and about equal for $\tan\beta\sim 2$ while the components of ${\bf S}_H$ have opposite sign, which considerably suppresses the coupling of $H$ to the $VV$ channels.

{\bf \emph{Decay rates}}. 
To determine the decay rates, we have used the program HDECAY~\cite{Hdecay} which gives the decay rates of the SM Higgs boson in exclusive decays, and have multiplied them by the appropriate factors~\footnote{In the usual notation, one defines: $S_{hd}=\cos\alpha$ and $S_{hu}=\sin\alpha$ and the couplings of $h$ and $H$ to $WW$ are $\cos (\beta-\alpha)$ and $\sin(\beta-\alpha)$ respectively.} taken from Tab. \ref{couplings} to \ref{couplings3}.

\begin{table}[hb]%
\label{tab:couplings}\centering%
\begin{tabular}{|lc|l|clc|l|c}
\hline%
 && $WW=ZZ$&& $t\bar t = c\bar c$ && $b\bar b =\tau^{+} \tau^{-}$\\
\hline
i=$h$, $H$ && $ \cos\beta \;S_{id} + \sin\beta \;S_{iu}$ && $ (\sin\beta)^{-1}S_{iu}$ && $(\cos\beta)^{-1}S_{id} $ \\
\hline
\end{tabular}
\caption{\label{couplings}Ratios of the couplings of $h$ and $H$ to the SM Higgs boson couplings, for different exclusive channels.}
\end{table}
\begin{table}[hb]%
\label{tab:couplings2}\centering%
\begin{tabular}{|cc|l|clc|l|c}
\hline%
 && $hZ$&& $t\bar t = c\bar c$ && $b\bar b =\tau^{+} \tau^{-}$\\
\hline
$A$ && $ \sin\beta \;S_{id} - \cos\beta \;S_{iu}$ && $ \cos\beta$ && $\sin\beta $ \\
\hline
\end{tabular}
\caption{\label{couplings2}Ratios of the couplings of $A$ to the SM Higgs boson couplings, for different exclusive channels.}
\end{table}

\begin{table}[hb]%
\label{tab:couplings3}\centering%
\begin{tabular}{|lc|l|cl|}
\hline%
 && $hW^+$&& $t\bar b = c\bar s=\nu_\tau\tau^{+}$\\
 \hline
 $H^+$ &&$\cos\beta\; S_{1u}-\sin\beta\;S_{id}$&&$-\sin\beta$\\
\hline
\end{tabular}
\caption{\label{couplings3}Ratios of the couplings of $H^+$ to the SM Higgs boson couplings, for different exclusive channels.
}
\end{table}
Decay rates in $\gamma\gamma$ are dominated by $WW$ and $t\bar t$ loops.  We take from Ref.~\cite{marciano} the separate SM loop amplitudes and rescale them with the appropriate couplings given in the previous Table.

{\bf \emph{Cross sections}}. 
The total production cross section for the Higgs particles is dominated by gluon-gluon fusion, which in turn is dominated  by the top quark loop. However, CMS searches also for  $\gamma \gamma$ events in the central region, accompanied by two, backward and forward hadron jets. In these conditions, Vector Boson Fusion (VBF) is favored but, given the CMS cuts, it is not $100\%$. 
The ATLAS analysis enhances vector boson fusion as well, but we will focus on the CMS set of cuts.

\begin{figure}[htb]
\begin{center}
\includegraphics[scale=.40]{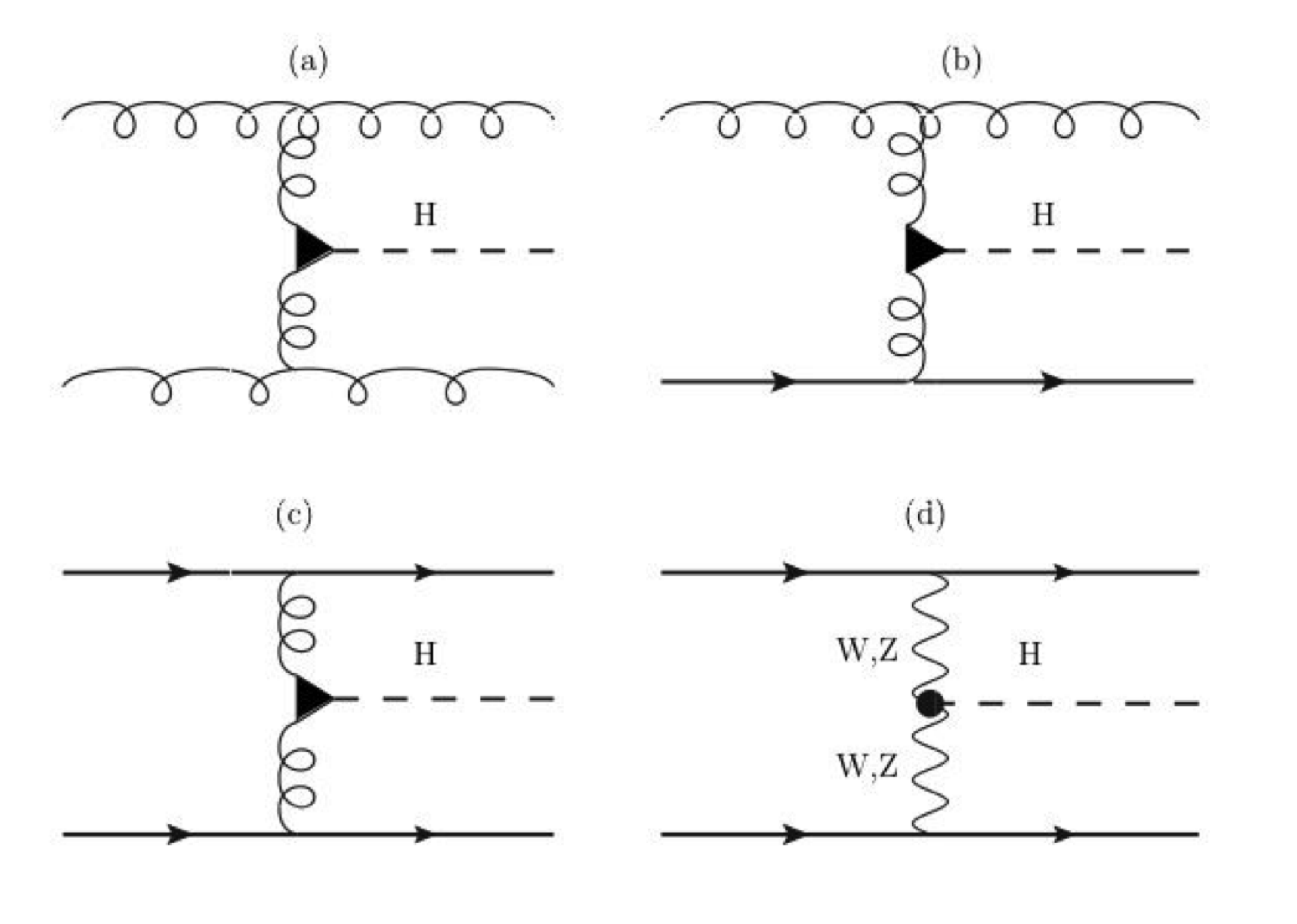}
\caption{
{\small 
Feynman diagrams contributing at parton level to the process $p+p\to H+2\;jets$. Full triangles: the gluon-gluon-Higgs effective vertex induced by the top loop, full dot: VV-Higgs vertex.} 
}
\label{allgraphs}
\end{center}
\end{figure}

We split the cross section for Higgs production accompanied by two jets in four different categories which we label as $gg$,  $gq$,  $qq$ and  VBF, see Fig. \ref{allgraphs}, diagrams (a), (b), (c) and (d), respectively. To compute them we use the libraries in~\cite{alpgen}.

We give in Tab. \ref{xsectionsh} the values of the SM cross sections for $h+2\; {\rm jets}$  ($M_h=125$ GeV) and   $H+2\; {\rm jets}$  ($M_H=320$ GeV), separately  for two cases:  $i)$~the {\it minimal cuts}, where we require only a maximum value of jets pseudo rapidity, $\eta_{\rm max}= 4.7$, and a minimum value of the jets transverse momentum, $(p_T)_{\rm min}=20$ GeV and $ii)$~{\it CMS-cuts}, where we enforce the cuts applied by CMS to the diphoton events of their  fifth category~\cite{bump}. The CMS requirements are:

\begin{itemize}
\item $E_T$ thresholds for the two jets of $30$ and $20$ GeV;
\item pseudorapidity separation between jets greater than $3.5$;
\item jet-jet invariant mass greater than $350$ GeV;
\item difference between the average pseudorapidity of the two jets and the pseudorapidity of the diphoton system (i.e. the Higgs boson) less than $2.5$.
\end{itemize}
The final requirement imposed by CMS is automatically enforced in Monte Carlo calculations.
 
 \begin{table}[hb]
\label{tab: xsectionsh}
\centering%
\begin{tabular}{|cc|clc||c|c|}
\hline
                  &         & $gg$     & $gq$        & $qq$       &  ${\rm VBF}$ \\
\hline 
 $h,\;M_h=125$ GeV & minimal cuts & $0.827$ & $ 0.674$ & $0.062$ & $0.826$ \\
                                         & CMS cuts       & $0.026$ & $0.056$ & $0.018$ &  $0.361$ \\
\hline
$H,\;M_H=320$ GeV & minimal cuts & $0.162$ & $ 0.137$ & $0.014$ &  $0.164$ \\
                                          & CMS cuts      & $0.005$ & $0.013$ & $0.005$ &  $0.102$ \\
\hline
\end{tabular}
\caption{Cross sections (pb) for: $p+p\to {\rm Higgs+2jets}$.}
\label{xsectionsh}
\end{table}
 
Our results show that, indeed, the CMS cuts favour considerably VBF over the gluonic channels, which, however survive the cuts  to a non negligible extent. To get a more quantitative estimate, we have also considered the inclusion of the so-called $K$-factors, that provide for a consistent part of the higher order corrections, not included in~\cite{alpgen}. For example we use the estimates given in Ref.~\cite{Huston}, assuming the numerical values:
\be
K_{gg} =1.72; \; K_{qq}= 1.23
\label{kfactors}
\ee
and compute the corrected cross sections by multiplying the values in Tab. \ref{xsectionsh} by the appropriate factors:
\bea
&&\sigma_{g\rm corr}=  K_{gg} \sigma_{gg}+ \sqrt{K_{gg}K_{qq}} \sigma_{gq}+ K_{qq}\sigma_{qq} \nonumber \\
&& \sigma_{{\rm VBF} \rm  corr}= K_{qq}\sigma_{{\rm VBF}}
\label{sigmacorr}
\eea
 \begin{table}[hb!]%
\label{tab: corrxsects}\centering%
\begin{tabular}{|cc|c|c|l|c|}
\hline
                                           &                          & $\sigma_{g\rm corr}$ & $\sigma_{{\rm VBF} \rm corr}$ & $\sigma_{\rm TOT \rm corr}$  \\
\hline
 $h$, $M_h=125$ GeV & minimal cuts   & $2.48$                     & $1.02$                        & $3.50$ \\
     &CMS cuts & $0.149$ & $0.444$ & $0.593$\\
 \hline
$H$, $M_H=320$ GeV  & minimal cuts &  $0.495$ & $ 0.202$ & $0.697 $ \\
    &CMS cuts &  $0.035$ & $0.125 $ & $ 0.160$ \\
\hline
\end{tabular}
\caption{Same as Tab. \ref{xsectionsh}, with cross sections rescaled by $K$-factors, see eq. (\ref{sigmacorr}). }
\label{corrxsects}
\end{table}
 Results for $h$ and $H$ are shown in Tab. \ref{corrxsects}.
For reference we have computed with the same libraries~\cite{alpgen} the total cross section for the SM inclusive Higgs production at LHC as a sum of the  cross sections for Higgs production with 0, 1 and 2 jets. Including the $K$-factors (\ref{kfactors}) and compared with more accurate calculations available. For $M_h=125$ GeV, we obtain:
\be 
\sigma (0+1+2~{\rm  jets}) \sim 17\; {\rm pb}
\ee
to be compared with the inclusive Higgs cross section at the same mass \cite{deflorian, anastasiou}:
\bea
&&\sigma({\rm inclusive}) = 19.49\; {\rm pb} \;({\rm dFG}) \nonumber \\
&&\sigma({\rm inclusive}) = 20.69\; {\rm pb}\; ({\rm ABHL})
\eea
 
 We find that, indeed, the CMS cuts favor considerably VBF over the gluon channels, which, however survive to a non negligible extent. In the case of $h$ we get 
 \begin{equation}
 \left(\sigma_{{\rm VBF}\rm corr}:\sigma_{g\rm corr}\right)_{h}\simeq 3:1
\end{equation} 
which agrees with what stated in Ref.~\cite{bump}. For $H$, at $m_H=320$~GeV:
\begin{equation}
 \left(\sigma_{{\rm VBF}\rm corr}:\sigma_{g\rm corr}\right)_{H}\simeq 3.5:1
\end{equation} 


Ratios of production cross sections to the SM ones can now be computed upon specifying the dominant mechanism:
\bea
&&R_{gg}(h)=\frac{\sigma(gg\to h)^{{\rm MSSM}} }{\sigma(gg\to h)^{{\rm SM}} }= \frac{S_{hu}^2}{ \sin^2\beta}\nonumber \\
&&R_{{\rm VBF}}(h)=\frac{\sigma(VV\to h)^{{\rm MSSM}} }{\sigma(VV\to h)^{{\rm SM}}}= ( \cos\beta \; S_{hd} + \sin\beta \; S_{hu})^2
\eea
and similarly for $H$.

{\bf \emph{Results}}. 
 In the numerical calculations, we take as a guiding values for $M_H$:
 \begin{itemize} 
  \item the value $M_H=320$ GeV, in correspondence to which an ``excess" can be seen in CMS data, albeit with small significance at about $1/3$ with respect to SM \cite{bump};
 \item the value $M_H=500$ GeV, to exemplify the behavior above the $t \bar t$ threshold. 
 \end{itemize}

\begin{figure}[htb]
\begin{minipage}[t]{80mm}
\includegraphics[height=6.0truecm, width=8truecm]{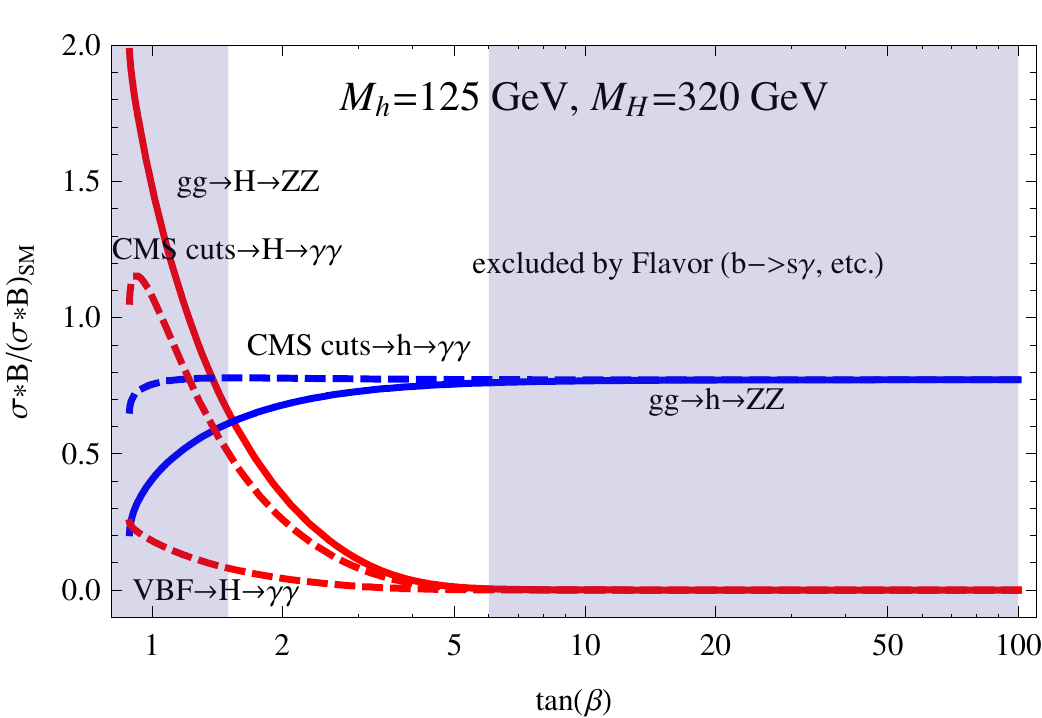}
\caption{\label{ZZplot} \footnotesize Values of $\frac{\sigma \times BR}{(\sigma\times BR)^{\rm SM}}$ in the $ZZ$ and $\gamma \gamma$ channels for $h$, $M_h=125$ GeV,  and $H$, $M_H=320$ GeV.  Gluon-gluon fusion is assumed for $ZZ$ and VBF, to the degree allowed by the CMS cuts for $\gamma\gamma$. The lower dashed curve $\mathrm{VBF}\to H\to \gamma\gamma$ is obtained by assuming $100\%$~VBF.
}
\end{minipage}
\hspace{\fill}
\begin{minipage}[t]{75mm}
\includegraphics[height=6.0truecm, width=8truecm]{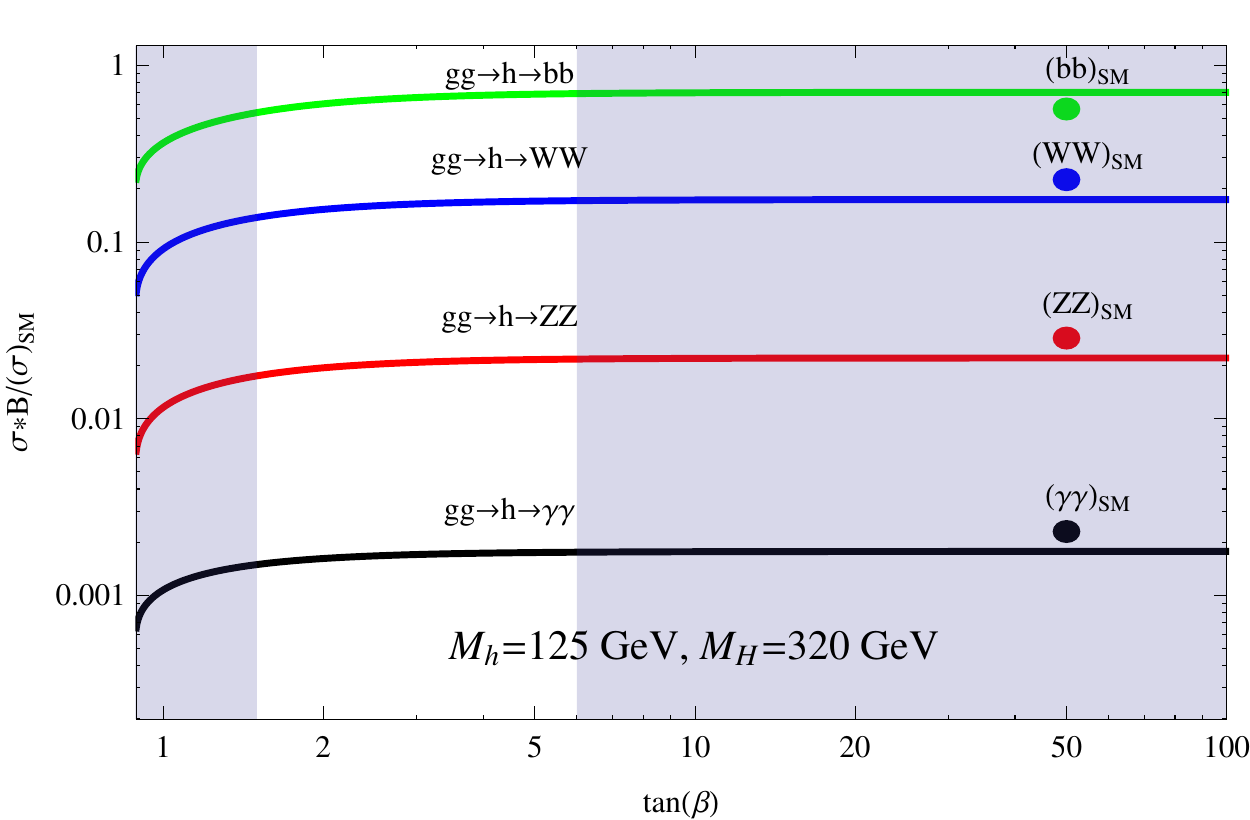}
\caption{\label{figh} \footnotesize Values of $\frac{\sigma \times BR}{\sigma^{\rm SM}}$ for h, $M_h=125$ GeV and $M_H=320$. For comparison, the same ratios in SM are shown by the dots in the right hand corner. }
\end{minipage}
\end{figure}

\begin{figure}[htb]
\begin{minipage}[t]{80mm}
\includegraphics[height=6.0truecm, width=8truecm]{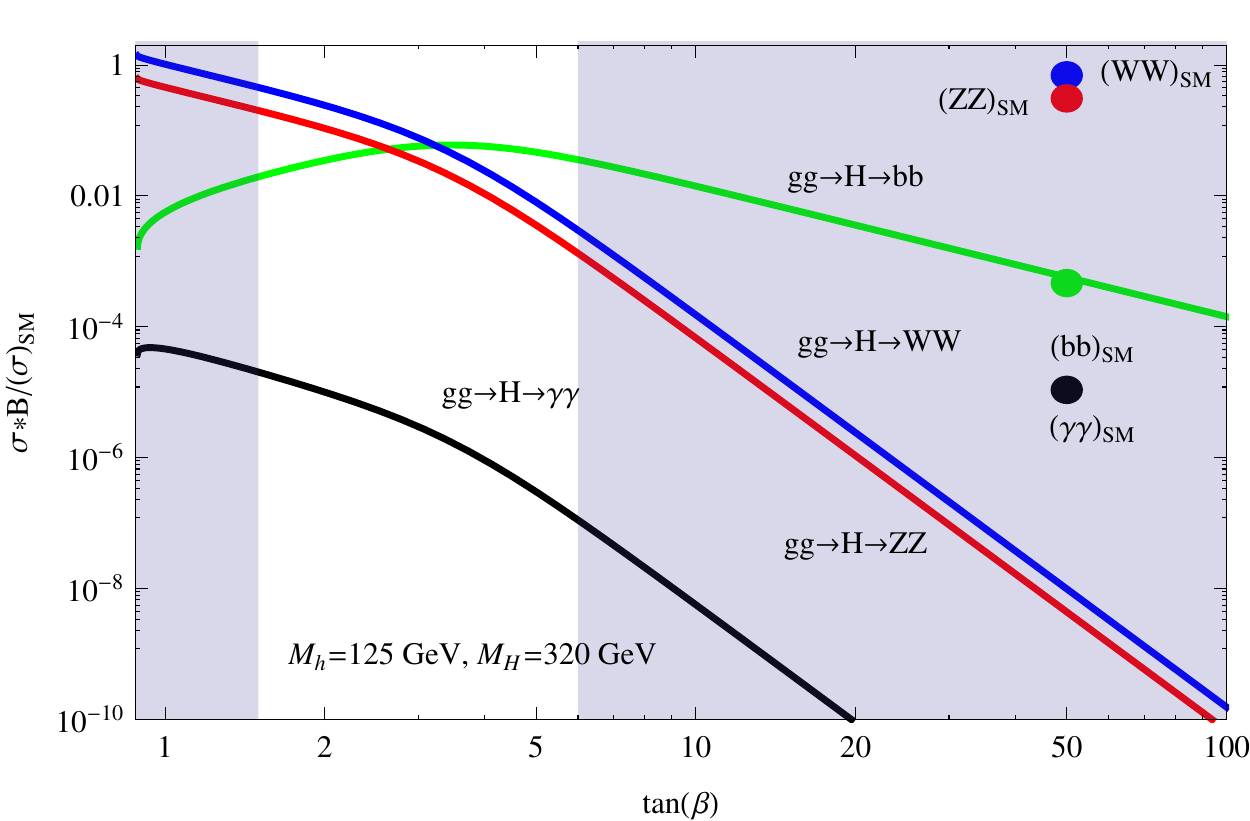}
\caption{\label{figH320} \footnotesize Values of $\frac{\sigma \times BR}{\sigma^{\rm SM}}$ for H, $M_H=320$ GeV. SM values shown in the right corner. }
\end{minipage}
\hspace{\fill}
\begin{minipage}[t]{75mm}
\includegraphics[height=6.0truecm, width=8truecm]{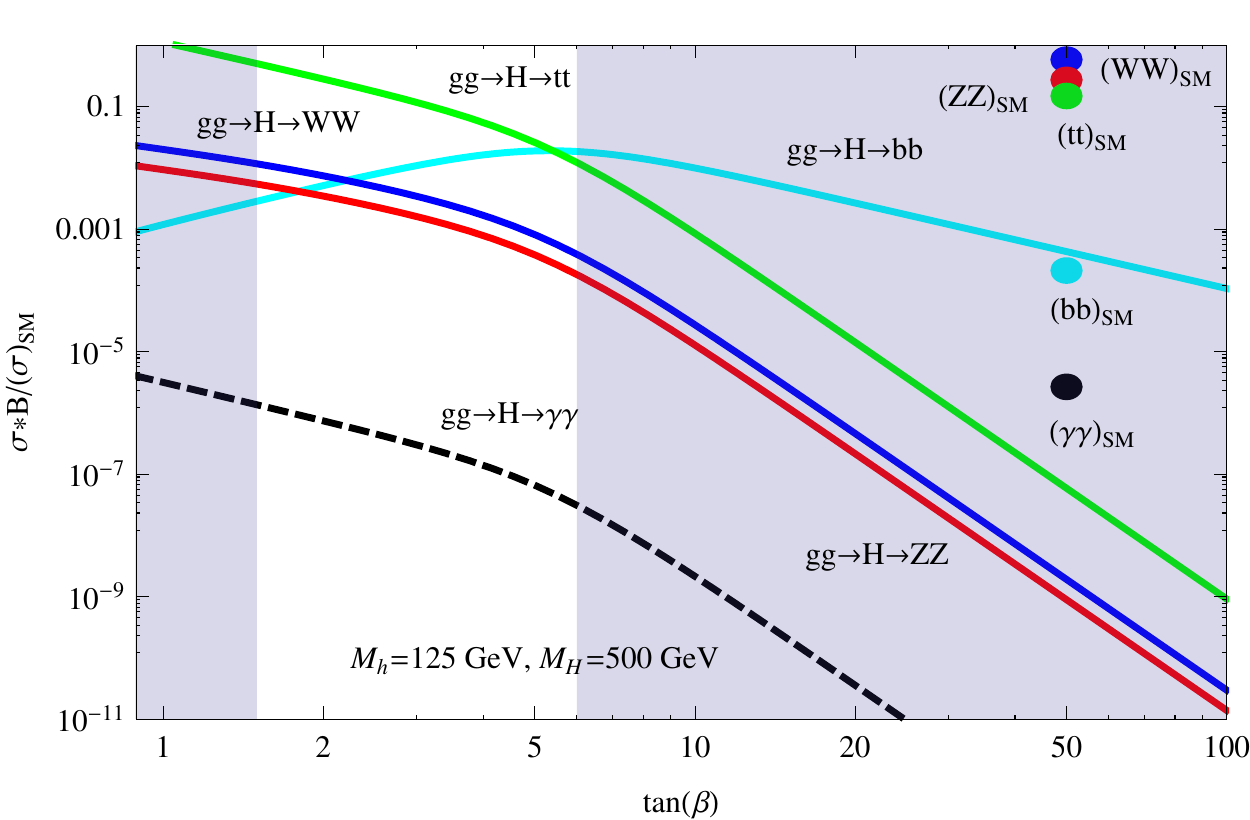}
\caption{\label{figH500} \footnotesize Values of $\frac{\sigma \times BR}{\sigma^{\rm SM}}$ for H, $M_H=500$ GeV. SM values shown in the right corner.}
\end{minipage}
\end{figure}

Results are summarized in
Figs. \ref{nasoverde}, \ref{pseudo}  and  \ref{ZZplot} to \ref{figH500}.  The shaded areas in the figures indicate 
the presently excluded regions of $\tan \beta$ corresponding to: $i)$ the region where  the $(\sin \beta)^{-1}$  factor would make the $t$ quark Yukawa coupling  to run out of the perturbative region before the GUT scale is reached~\cite{Yeghian:1999kr} $ii)$ the limit on $\tan \beta$ derived~\cite{Kaneb, nmah} from the non observation of FCNC decays of $B$ mesons such as $B_{s}\to \mu^+\mu^-$~\cite{lhcbmm}. 

\paragraph{{\bf $h$(125).}}  For $M_H=320$~GeV, we find that the  light Higgs boson $h$ behaves rather closely like the SM  Higgs boson, in both $ZZ$ and $\gamma \gamma$ channels, Fig. \ref{ZZplot}. For $M_H=500$~GeV the decoupling limit is reached and $h$ is essentially a SM Higgs boson.
We adopt the gluon-gluon fusion mechanism for the $ZZ$ channel and for $\gamma \gamma$ we use the cross sections in Tab. \ref{corrxsects} with CMS cuts, with the gluon fusion and  vector boson fusion appropriately rescaled with SUSY couplings.  


We show in Fig.~\ref{figh} the values of $\frac{\sigma \times BR}{\sigma^{{SM}}}$ for SUSY $h$, $M_h=125$ GeV and  $M_H=320$~GeV.  The value of $\sigma\times BR$ is obtained by multiplying by the SM cross section for a Higgs boson of the same mass.
For comparison, the same ratios in SM are shown by the dots in the right hand corner, for the different channels.  The $b \bar b$ branching ratio is $\sim 1.5$ larger than the SM one, due to the factor $(\cos \beta)^{-1}$ in the Yukawa coupling.

\paragraph{{\bf $H$(320/500).}}The vector boson fusion $VV\to H\to ZZ$ gives a negligible result and we are left with  $gg$ fusion as the production mechanism. For $\gamma\gamma$ we use the results of  Tab. \ref{corrxsects} with CMS cuts. Not surprisingly, a large difference is made by being below or above the top quark threshold. 

For $M_H=320$ GeV, the ratios of $\sigma \times BR(H\to ZZ)$ and $\sigma \times BR(H\to \gamma \gamma)$in SUSY over the same in SM drop very fast at the increase of $\tan\beta$, see Fig. \ref{ZZplot}. In a window around $\tan \beta =2$, $H$ would appear as a subdominant companion of $h$ in {\it both} $VV$ and $\gamma \gamma$ channels.  Note that the assumption of pure VBF would give a very small result for $\gamma \gamma$, see Fig.\ref{ZZplot}.  For $\tan\beta \geq 4$, the $VV$ and $\gamma \gamma$ decay branching ratios drop and the $b \bar b$ channel takes over, due to the  $(\cos \beta)^{-1}$ factor, Fig. \ref{figH320}.  

Taking seriously the bump at $M_H=320$~GeV, we get from Fig.  \ref{ZZplot}:
\be
\frac{\sigma\times BR(H\to ZZ)^{{\rm MSSM}}}{(\sigma\times BR)^{{\rm SM}}}\sim 0.3 \;\;\; {\rm at}\;\;\; \tan\beta \sim 2 
 \ee
 and, in correspondence, see also Figs. \ref{nasoverde} and \ref{pseudo}and Eq. (\ref{mcharged}): 
 \bea
&&\frac{\sigma\times BR(h\to\gamma\gamma)^{{\rm MSSM}}}{(\sigma\times BR)^{{\rm SM}}}\sim 0.8\\
&& \frac{\sigma\times BR(H\to\gamma\gamma)^{{\rm MSSM}}}{(\sigma\times BR)^{{\rm SM}}}\sim 0.26 \;\;(\tan \beta =2, \;\;{\rm CMS\;cuts})\\
&&M_A = 310\; {\rm GeV};\; M_{H^\pm} = 320\; {\rm GeV}\label{mahp}\\
&& \sqrt{M_{\tilde t_R}M_{\tilde t_L}}=3.9\;{\rm TeV} \label{mstop}
 \eea

For $M_H=500$ GeV, the $t\bar t$ channel is dominating up to $\tan\beta \sim 5$ and the observation of $H$ is related to the ability to detect top quarks. For $\tan \beta\geq 2$ the $b \bar b$ channel is also significant.

The ratios $\frac{\sigma \times BR}{\sigma^{\rm SM}}$  give a measure of the visibility of the different channels for $H$. We plot these ratios for $H$  below and above the $t\bar t$ threshold in Figs. \ref{figH320} and \ref{figH500}. 


\paragraph{{\bf $A$ and $H^\pm$.}} Given the masses of the CP-odd neutral $A$ and of the charged $H^\pm$ bosons, given in~(\ref{mahp}),  on can foresee the dominant decays~\cite{Hdecay}  $A\to b\bar b,\;  \tau \bar \tau,\;  Z h$, with branching ratios $(0.43,\; 0.06,\; 0.36)$ respectively and $H^+\to t\bar b, \; h W^+  ~(0.99,\;  0.007)$.

{\bf \emph{Conclusions}}. A clear prediction of Supersymmetry is the presence of two Higgs field doublets, one coupled to $u$ and the other to $d$ quarks. If the $125$ GeV signal is confirmed, the next thing to look for is the presence of a secondary line in $VV$, $\gamma \gamma$ and $b \bar b$. We have shown that these signals are viable for $M_H$ below the $t \bar t$ threshold and in the rather narrow region of $\tan\beta$ allowed at present. An $H$ at $320$ GeV seen in $ZZ$ and $\gamma \gamma$ would fit very well into MSSM with $\tan\beta\sim2$ and this calls for close scrutiny of this region.
Clear signatures for the CP-odd $A$ and the charged $H^\pm$ are the $b\bar b$ and $t\bar b$ decays respectively. In all cases we have considered, a scalar top mass around $4$~TeV is predicted. 

\vskip0.7cm

{\bf \emph{Acknowledgements}}. We are grateful to R. Barbieri, D. Del Re, C. Dionisi, F. Gianotti, G. Organtini and G. Tonelli for illuminating discussions.

\end{document}